\documentclass[a4paper]{article}

\usepackage{INTERSPEECH2022}
\usepackage{xcolor}
\usepackage{balance}
\usepackage{hyperref}

\title{PEAF: Learnable Power Efficient Analog Acoustic Features \\for Audio Recognition}%

\name{Boris Bergsma$^{\dagger}$$^{\star}$ \qquad Minhao Yang  $^{\diamond}$ \qquad Milos Cernak$^{\dagger}$ }

\address{$^{\star}$ École Polytechnique Fédérale de Lausanne (EPFL), Switzerland  \\
      $^{\dagger}$ Logitech Europe S.A., Lausanne, Switzerland \\
       $^{\diamond}$ Reexen Technology, Zürich, Switzerland
}
\email{bergsma.boris@gmail.com, yang@reexen.com, milos.cernak@ieee.org}

\begin{document}

\maketitle
\begin{abstract}
% Minhao
At the end of Moore’s law, new computing paradigms are required to prolong the battery life of wearable and IoT smart audio devices. Theoretical analysis and physical validation have shown that analog signal processing (ASP) can be more power-efficient than its digital counterpart in the realm of low-to-medium signal-to-noise ratio applications. In addition, ASP allows a direct interface with an analog microphone without a power-hungry analog-to-digital converter. Here, we present power-efficient analog acoustic features (PEAF) that are validated by fabricated CMOS chips for running audio recognition. Linear, non-linear, and learnable PEAF variants are evaluated on two speech processing tasks that are demanded in many battery-operated devices: wake word detection (WWD) and keyword spotting (KWS). Compared to digital acoustic features, higher power efficiency with competitive classification accuracy can be obtained. A novel theoretical framework based on information theory is established to analyze the information flow in each individual stage of the feature extraction pipeline. The analysis identifies the information bottleneck and helps improve the KWS accuracy by up to 7\%. This work may pave the way to building more power-efficient smart audio devices with best-in-class inference performance.
\end{abstract}

\section{Introduction}
\label{sec:Intro}

% Minhao
The emerging and expanding wearable and IoT market requires extremely power-efficient electronics for prolonged continuous operation with tiny batteries and/or energy harvesting. Audio devices like wireless stereo earbuds and active noise canceling earbuds, and other wearables like watches and glasses, can benefit from smart acoustic interfaces for gesture-free voice control. Such interface typically comprises always-on functionalities such as voice activity detection (VAD), wake-word detection (WWD), keyword spotting (KWS), etc. As always-on functionalities, the power efficiency of their hardware implementation dramatically affects the overall system battery life, especially when the power-hungry blocks triggered by them are heavily duty-cycled.

%With the new generation of intelligent devices in our daily lives, the user-device relationship has drastically changed in the last decade. Gesture-free instant communication is an appealing means of communication for users and manufacturers, especially when no touch interface is possible or restrictive. This kind of always-on listening platform would benefit from on-device efficient inference, which could primarily alleviate the concerns for energy consumption from data transmission and data privacy. One promising method is to use analog processing for acoustic feature extraction~\cite{linear,nonlinear}. The power consumption advantage of analog processing is twofold. Firstly, the analog processing itself is more efficient than the digital-based on the Mel-frequency cepstral coefficient (MFCC)~\cite{mfcc-energy} in the case of low-to-medium signal-to-noise ratio requirement~\cite{hosticka1985performance}. Secondly, the analog processing can directly interface with an analog MEMS microphone without needing a power-consuming high-precision Analog-to-Digital Converter (ADC). A state-of-the-art 16-bit ADC~\cite{chandrakumar201815} can consume ten times more power than the MFCC computation~\cite{mfcc-energy} itself.

With 3nm process looking to be in mass production in 2022/2023, the 50 years of silicon scaling will end in this decade. Not only the active power reduction associated with transistor feature size shrinkage is soon finished, but also the severe leakage problem in advanced technology nodes largely compromises the power efficiency. Other than the dominant digital signal processing (DSP), new computing paradigms are being actively explored, and analog signal processing (ASP) is one promising alternative. The power advantage of ASP is twofold. Firstly, theoretical analysis~\cite{hosticka1985performance} and experimental validation~\cite{linear,nonlinear} show that ASP is more power efficient than its digital counterpart~\cite{raychowdhury20132,mfcc-energy} implementing algorithms like Mel-frequency cepstral coefficient (MFCC) when low-to-medium signal-to-noise ratio processing is sufficient. Secondly, ASP can directly interface with an analog MEMS microphone~\cite{linear,nonlinear}, avoiding a power-consuming high-precision analog-to-digital converter (ADC)~\cite{chandrakumar201815}. Recent chip implementations of ASP-based acoustic feature extraction focusing on the VAD functionality~\cite{linear,nonlinear} have showed the extreme low power consumption from nanowatts to microwatts with inference accuracy comparable to that of MFCC-based features.

%Neuromorphic computing based on spiking neural network-based audio recognition has also been proven to be power efficient~\cite{Dellaferrera_2020,martinelli2020spiking}. Unfortunately, missing chips postpone the usability of these methods. This work thus focuses on analog features with already fabricated analog chips for feature extraction and digital chips for audio recognition (classification).

%Recent works on analog features focus only on voice activity detection (VAD) with the power consumption in the range of nano to microwatts~\cite{linear,nonlinear} and inference accuracy comparable to that of MFCC-based features. 

\begin{figure}[htb]
  \centering
  \includegraphics[width=\linewidth]{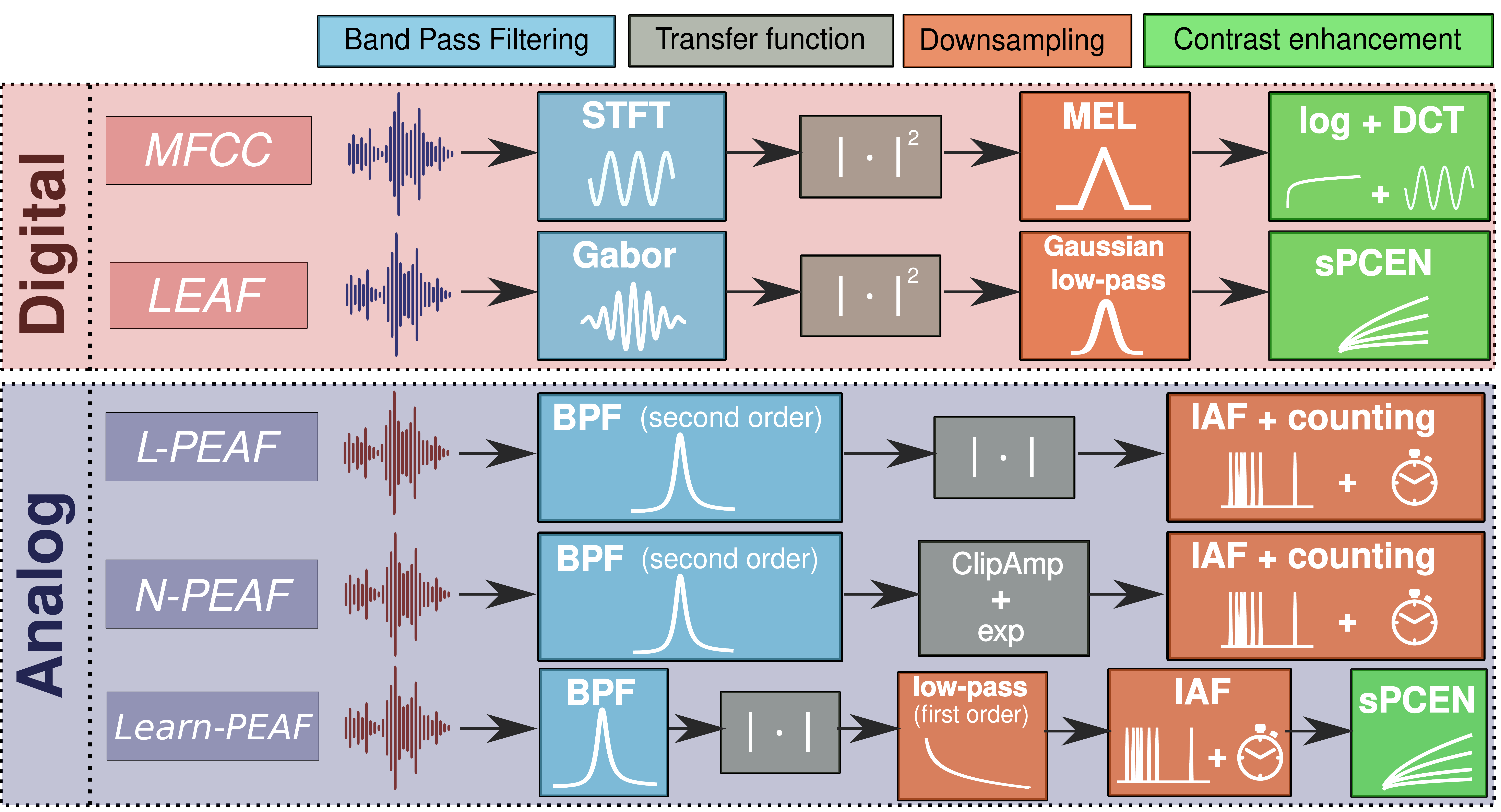}
\caption{Scheme of all the compared features in this paper. STFT : Short Term Fourier Transform, DCT: Discrete Cosine Transform, PCEN: Per Channel Energy Normalisation, BPF: BandPass Filters, IAF: Integrate And Fire encoder, L-: Linear, N-: Non-linear, PEAF: Power Efficient Analog Features.}
\label{fig:Analog_schemeL}
\end{figure}

In this work, linear~\cite{linear} (L-PEAF), non-linear~\cite{nonlinear} (N-PEAF), and proposed learnable power-efficient analog acoustic features (Learn-PEAF) are evaluated on WWD and KWS, which are highly sought-after functionalities in many wearable and IoT devices. Given that the PEAF variants have spike output in the feature extraction pipeline, spiking neural networks~\cite{Dellaferrera_2020,martinelli2020spiking} seem to be the natural candidate as the classifier. But we use artificial neural networks because of the state-of-the-art energy efficiency in their hardware implementations. Inspired by a learnable audio frontend (LEAF)~\cite{leaf} intended for improving DSP-based feature extraction, we devise a novel analysis method based on information theory and use it for the learnable PEAF design. Our experimental validation on both WWD and KWS shows improved inference performance with appreciable power reduction. Fig. \ref{fig:Analog_schemeL}. shows all the considered features in this work.

%This work presents power-efficient analog acoustic features (PEAF) that are complemented by conventional digital chips for running audio recognition. Linear, non-linear, and learnable PEAF variants are evaluated on two speech processing tasks that are demanded in many battery-operated devices: wake work detection~\cite{tang2020howl} and keyword spotting (KWS)~\cite{warden2018speech}. Wake word detection is a simple binary classification task, while KWS is a more complex 35 class classification problem.

%Minhao
%Moreover, this paper devises an information theory-based analysis of compared feature representations to find the bottleneck operations in the computation of features. All analog and MFCC representations are compared to LEAF~\cite{leaf}, learnable feature extraction that outperforms the MFCC in multiple tasks while using a bandpass filter bank. Its processing pipeline is similar to the analog ones, and it allows to locate the information bottleneck in the system to help further improve the analog audio representation. The theoretical claims are then validated by concrete experiments.

\section{Methods}
\label{sec:TH}
Digital handcrafted (MFCCs) and learnable (LEAFs) features are compared to linear (L-PEAF), non-linear (N-PEAF), and learnable (Learn-PEAF) analog features.
 
\subsection{(Reference) digital features}

%Due to its popularity the digital feature used as reference is the MFCC. \cite{mfcc-energy} is used as a reference, having as parameters: 20 Mel-frequency bands and 10 coefficients.   
%LEAF ~\cite{leaf} is a method that outperforms MFCC in multiple task. It follows a similar computation pipeline as our PEAF features, making for an interesting point of comparison. 

Due to our ambition to compare the performance and power efficiency of features on fabricated chips, we have chosen MFCC reference implementation~\cite{mfcc-energy} with twenty Mel-frequency bands and ten coefficients. LEAF is a method that outperforms MFCC in multiple tasks, but its neural network based implementation~\cite{leaf} has not provided power measurements and we thus excluded it from comparison. Instead, we used the LEAF representation in information theory-based analysis (Sections ~\ref{sec:INF} and \ref{sec:analog_improvement}) to identify suboptimal stages in analog feature computation.

\subsection{Analog features}

%The schema of analog feature computation that roughly follows the human cochlea processing are shown in Fig.~ \ref{fig:Analog_schemeL}. The computation comprises three basic operations: 2nd-order bandpass filtering, an activation function, and spike generation based on Integrate-and-Fire (IAF) encoding. The IAF functions like a biological neuron, accumulating a certain amount of electrical charge and releasing a voltage spike after reaching a certain threshold. We call the analog feature extraction with a absolute value activation function ``L-PEAF``, given its piece-wise linear nature and the linearity requirement on circuit implementation~\cite{linear}, and the one with a clipped exponential activation function ``N-PEAF``, which can be implemented with a limiting amplifier and a single voltage-to-current conversion transistor~\cite{nonlinear}. The output of the IAF encoder can be directly used for spiking neural networks~~\cite{Dellaferrera_2020} or counted in frames and used as a 2D feature like the MFCC. PEAF is our proposed improvement to the actual L-PEAF architecture. It has an additional first order low pass filter (per channel) before the IAF and a PCEN step before entering the digital classifier. All the PEAF features have build in and fixed parameters, for example the center frequency of the band pass filters. Our PEAF implementation was coded using PyTorch, allowing to optimise its parameters in conjunction with the deep learning classifier. Its parameters are optimized on Keyword Spotting with LeNet Classifier \textbf{(B)} (sec. \ref{sec:CNN}). 

Fig.~ \ref{fig:Analog_schemeL} shows the schema of analog feature computation that follows the human cochlea processing. The computation comprises three basic operations: 2nd-order bandpass filtering, an activation function, and spike generation based on Integrate-and-Fire (IAF) encoding. The IAF acts as a biological neuron, accumulating a certain amount of electrical charge and releasing a voltage spike after reaching a certain threshold. We call the analog feature extraction with an absolute value activation function ``L-PEAF``, given its piece-wise linear nature and the linearity requirement on circuit implementation~\cite{linear}. On the contrary, a version with a clipped exponential activation function is tagged ``N-PEAF``, which can be implemented with a limiting amplifier and a single voltage-to-current conversion transistor~\cite{nonlinear}. The output of the IAF encoder can be directly used for spiking neural networks~\cite{Dellaferrera_2020} or counted in frames and used as a 2D feature like the MFCC.

Learnable PEAF, named ``Learn-PEAF``, is our proposed improvement to the actual L-PEAF schema. It has an additional first-order low pass filter (per channel) before the IAF and a PCEN step before entering the digital classifier. All the PEAF features have built-in and fixed parameters, for example, the center frequency of the bandpass filters. The proposed Learn-PEAF feature schema was simulated using PyTorch. This allows to automatically differentiate the PEAF frontend, and the gradient can be back propagated through the classifier and frontend, allowing to optimise PEAF's parameters. 

\subsection{Digital deep learning classification}\label{sec:CNN}

\begin{figure}[htb]
  \centering
  \includegraphics[width=0.85\linewidth]{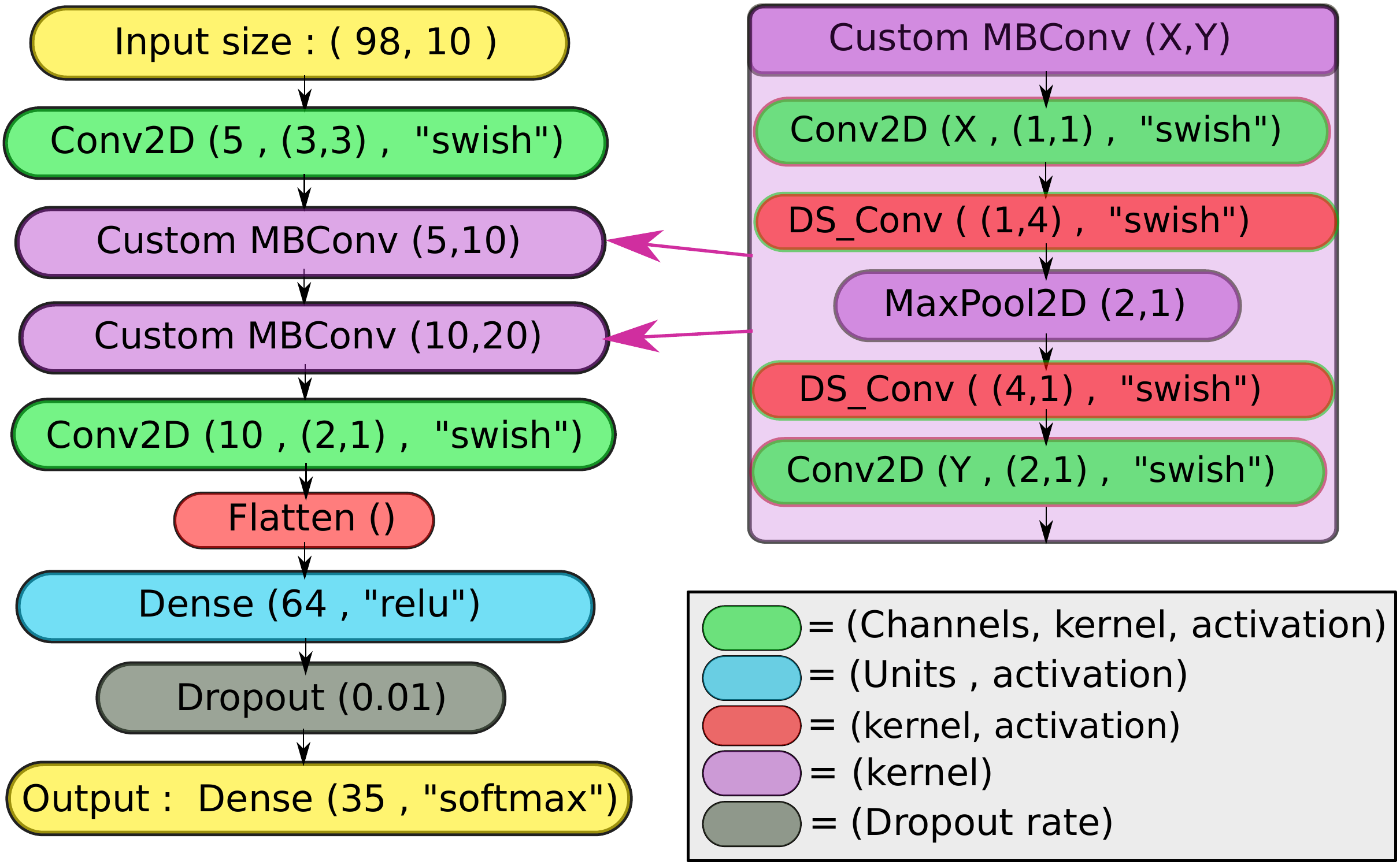}
\caption{Scheme of the Mini-EfficientNet classifier \textbf{(A)}.}
\label{fig:Classifiers}
\end{figure}

We evaluate the different features on multiple classifiers for scaling analysis and robust results. The employed classifiers have different neural network architectures with growing classification capacities: 
\textbf{A)} a small version of EfficientNetV2 \cite{Eff-V2} (Fig. \ref{fig:Classifiers}), \textbf{B)} LeNet-5~\cite{726791}, \textbf{C)} a depthwise separable CNN (DS-CNN) ~\cite{howard2019searching} and \textbf{D)} EfficientNetB0~\cite{tan2020efficientnet}. All the classifiers are trained using Adam optimization~\cite{kingma2014adam} with default learning rate, and SpecAugment \cite{Spec-augment} is used as data augmentation. A scheduler reduces the learning rate geometrically by e-0.01 after 100 epochs for the small neural networks (\textbf{A},\textbf{B}) and 50 for the large ones (\textbf{C},\textbf{D}).

\subsection{Power consumption estimation}

%Minhao
The power consumption of the L-PEAF and the N-PEAF can be estimated from the fabricated chips~\cite{linear,nonlinear}. The power consumption of the digital MFCC features can be estimated from~\cite{mfcc-energy}, plus an ADC’s contribution~\cite{chandrakumar201815}. The power consumption of the Learn-PEAF is obtained by estimating that the 1st-order lowpass filter consumes the same as the bandpass filters (16\% of the L-PEAF~\cite{linear}), and a possible simple DSP-based PCEN adds another 10\%. The power consumption estimates are summarized in Table~\ref{tabpow-features}. 

\begin{table}[ht!]
\caption{Power consumption of the different features.}
\vspace{4.5pt}
\centering
\begin{tabular}{cc}
\toprule
Feature & Power\\
\midrule
 MFCC \cite{mfcc-energy} + ADC \cite{chandrakumar201815}  & $0.34 + 7.2 = 7.5 \ \rm \mu W$\\
 N-PEAF \cite{nonlinear}             & $0.072 \ \rm \mu W$\\
 L-PEAF \cite{linear}                 & $0.38 \ \rm \mu W$\\
%  Learn-PEAF             & $0.48 \ \rm \mu W$\\
\bottomrule
\end{tabular}
\vspace{-9pt}
\label{tabpow-features}
\end{table}

To estimate a lower bound of the power consumption of the classifiers, the energy efficiency from one state-of-the-art neural network processor $E_{eff} = 36.5 \cdot 10^{12}\  \rm OPS / W$ is used~\cite{tu202228nm}. Table~\ref{tabpow-class} summarizes the power consumption of the considered classifiers. The classifiers’ power is calculated by equation~\ref{lab::Pow_decomp}, where $N_{OPS}$ is the number of operations done by the neural network model, $FR$ is the frame rate. We choose $FR=10fps$ for WWD and $FR=30fps$ for KWS.

\begin{equation}\label{lab::Pow_decomp}
    P_{tot}  = P_{feat} + P_{class} = P_{feat} 
    + \frac{N_{OPS} \times FR }{E_{eff}}
\end{equation}

\begin{table}[ht!]
\caption{Power consumption estimates of the classifiers for the MFCC with 10 coefficients.}
\vspace{4.5pt}
\centering
\begin{tabular}{lcc}
\toprule
Classifier & KWS Power  & WWD Power\\
\midrule
\textbf{A)}    Mini-EfficientNet (Fig. \ref{fig:Classifiers})        & $0.097 \ \rm \mu W$  & $0.079 \ \rm \mu W$ \\
\textbf{B)}     LeNet-5 \cite{726791}  & $0.34 \ \rm \mu W$   & $0.30 \ \rm \mu W$ \\
\textbf{C)}     DS-CNN \cite{howard2019searching}  & $8.3 \ \rm \mu W$   & $5.4 \ \rm \mu W$ \\
\textbf{D)}     EfficientNet \cite{tan2020efficientnet} & $48 \ \rm \mu W$    & - \\
\bottomrule
\end{tabular}
\vspace{-9pt}
\label{tabpow-class}
\end{table}

%Total power of audio recognition is given by Eq. \ref{lab::Pow_decomp}.

\subsection{Information theory} \label{sec:INF}

Information theory is used to compare the different features to identify the bottleneck operation in the feature computation. The tool used is the Shannon Entropy~\cite{Shanon_entropy}:
\begin{equation}
  S_{Shannon} = - \sum_{i=0}^N p_i \log{(p_i)}
\end{equation}
with $p_i$ the probability of the data distribution. The estimation of the data distribution in the features can be done in multiple ways. To retain the spatial information of the features this work encodes spatial information by giving each data point a unique identifier (an integer) in a second dimension. This two-dimensional object is defined as : 
\begin{equation}
 \textrm{Encoded feature} =  ( \textrm{Flattenedfeature}^T ,\  \textrm{linspace(1,N)}^T )
\end{equation}
with ``data-point value`` and ``spatial label`` dimensions. %This two dimensional object is then fit with a two-dimensional histogram to obtain an estimation of the $p_i$'s. This methods should be seen as a test, there are currently no best way to extract entropy for matrices without probability priors. The histogram computation would also require more research, with the number of bins in the histogram greatly influencing results, they were set manually in this test to obtain a smooth histogram. 
This two-dimensional object is then fitted with a two-dimensional histogram to obtain an estimation of the $p_i$'s. This method should be seen as an approximation, as there is currently no universally acceptable way to extract entropy for matrices without probability priors. The proposed method of histogram computation also requires more research; some initial exploration is presented in~\cite{Bergsma-thesis}. 
%improvement could be done by not relying on existing libraries to compute the histograms, or computing distributions on multiple features at once.

\section{Experimental Results}

In all our experiments, digital MFCCs with various digital classifiers and known overall power consumption was the baseline. We evaluated analog PEAF with the same digital classifiers.

\subsection{Feature analysis and PEAF validation}\label{sec:analog_improvement}

%To improve the analog features, we first compared them to the MFCCs and LEAF features to understand where difference arise. The idea is to know the causes of the loss in accuracy of the analog features on keyword spotting. A thousand samples with equal class repartition are used for this analysis. The features are separated into their computation steps, and the entropy is computed as described in section \ref{sec:INF}. The average of the entropy on all the samples is then plotted with its standard deviation in Fig.\ref{fig:inf_leaf}.

First, we applied information theory to gain an insight into information flow across feature computation. We hypothesized that comparing the information flows from the MFCCs, LEAFs, and PEAFs could identify ``information bottleneck``, and inform the learnable PEAF which steps could benefit from data-driven optimization.

\begin{figure}[htb]
  \centering
  \includegraphics[width=\linewidth]{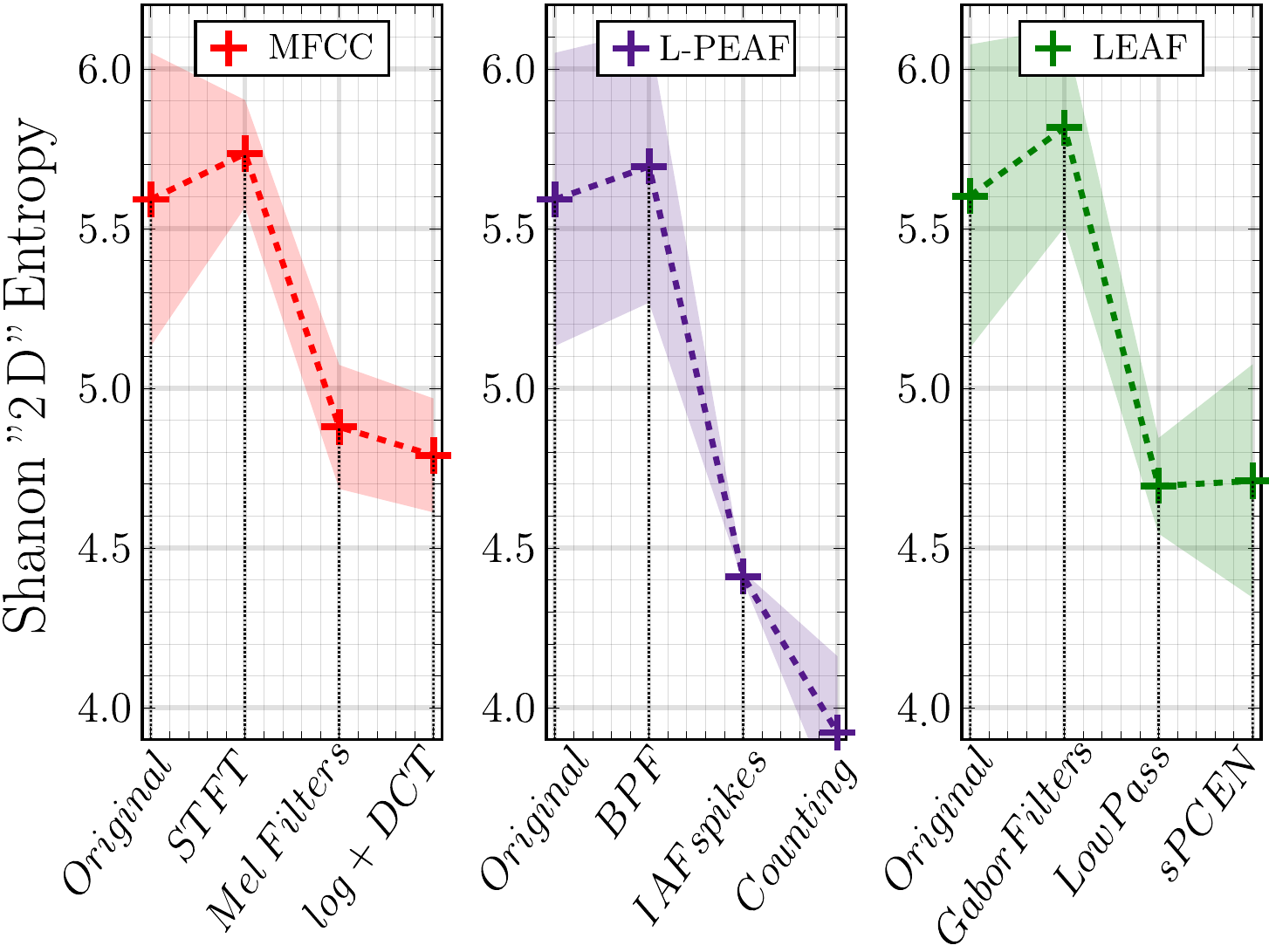}
\caption{Comparison of the features  at each computation step using information theory on KWS.}
\label{fig:inf_leaf}
\end{figure}

A thousand random samples from the Speech Command dataset V2~\cite{warden2018speech} with equal class repartition are used for this analysis. The features are separated into their computation steps, and the entropy is computed as described in section \ref{sec:INF}. The average of the entropy on all the samples is then plotted with its standard deviation in Fig.\ref{fig:inf_leaf}.

%The main hypothesis that can be extracted out of Fig.\ref{fig:inf_leaf} is that the last two steps of the computation of L-PEAF are the information bottleneck (more details in sec.\ref{sec:disc}). To confirm theses we tested adjustments in the analog feature computation by replacing them with LEAF's counterparts. The IAF was replaced with the Gaussian low pass filters of LEAF. Then, we added PCEN as the last step of the computation.

The obvious conclusion that can be made out of Fig.\ref{fig:inf_leaf} that the last two steps of the computation of analog features (L-PEAF) are the information bottleneck (more details in sec.\ref{sec:disc}). We tested adjustments in the analog feature computation by replacing them with LEAF's counterparts to validate it. The IAF was replaced with the Gaussian low pass filters of LEAF. Then, we added PCEN as the last step of the computation.
% Since we simulated PEAF computation in PyTorch, the parameters of the analog extractor were also able to be optimized using gradient descent. 
The results in Tab.~\ref{tab:inf_test} confirm that Learn-PEAFs outperform L-PEAF with the fixed parameters and also MFCCs. We validated the design of Learn-PEAFs and used the novel proposed features in consecutive experiments.
%PEAF is then added to both testing suites on Fig. \ref{fig:WWD_energy} and \ref{fig:KWS_Res} to even further validate the hypothesis. 
% Since the results of the improved analog were encouraging, we decided to test a proof of concept new analog extractor that could be implemented using only analog processes. Since the improvements seemed to arise from the low pass filtering and the addition of the PCEN, both of them were added to the processing pipeline in the form of a simple RC first order low pass filter and PCEN can be added digitally. A final scheme can is displayed in Fig. \ref{fig:Analog_schemeL}. 

\begin{table}[ht!]
\caption{Validation using Learn-PEAF with LeNet \textbf{(B)} classifier on KWS. For every row, one computation step of L-PEAF was replaced sequentially by the LEAF equivalent.}
\vspace{4.5pt}
\centering
\begin{tabular}{ccc}
\toprule
Parameters : & Fixed & Optimized\\
\midrule
L-PEAF (baseline) & 79.1\% & - \\
IAF $\rightarrow$ Gaussian Low Pass & 81.7\% & 82.8\%\\
Adding sPCEN & 82.6\% & \textbf{85.1\%}\\
% BPF $\rightarrow$ Gabor filters & - & 85.0\%\\ (Boris : I removed it, not really relevant in the discussion )
\midrule
Learn-PEAF & - & \textbf{86.4}\%\\
MFCC & 84.8\% & - \\
\bottomrule
\end{tabular}
\vspace{-9pt}
\label{tab:inf_test}
\end{table}

\subsection{Feature comparison for KWS task}

\begin{figure}[htb]
  \centering
  \includegraphics[width=\linewidth]{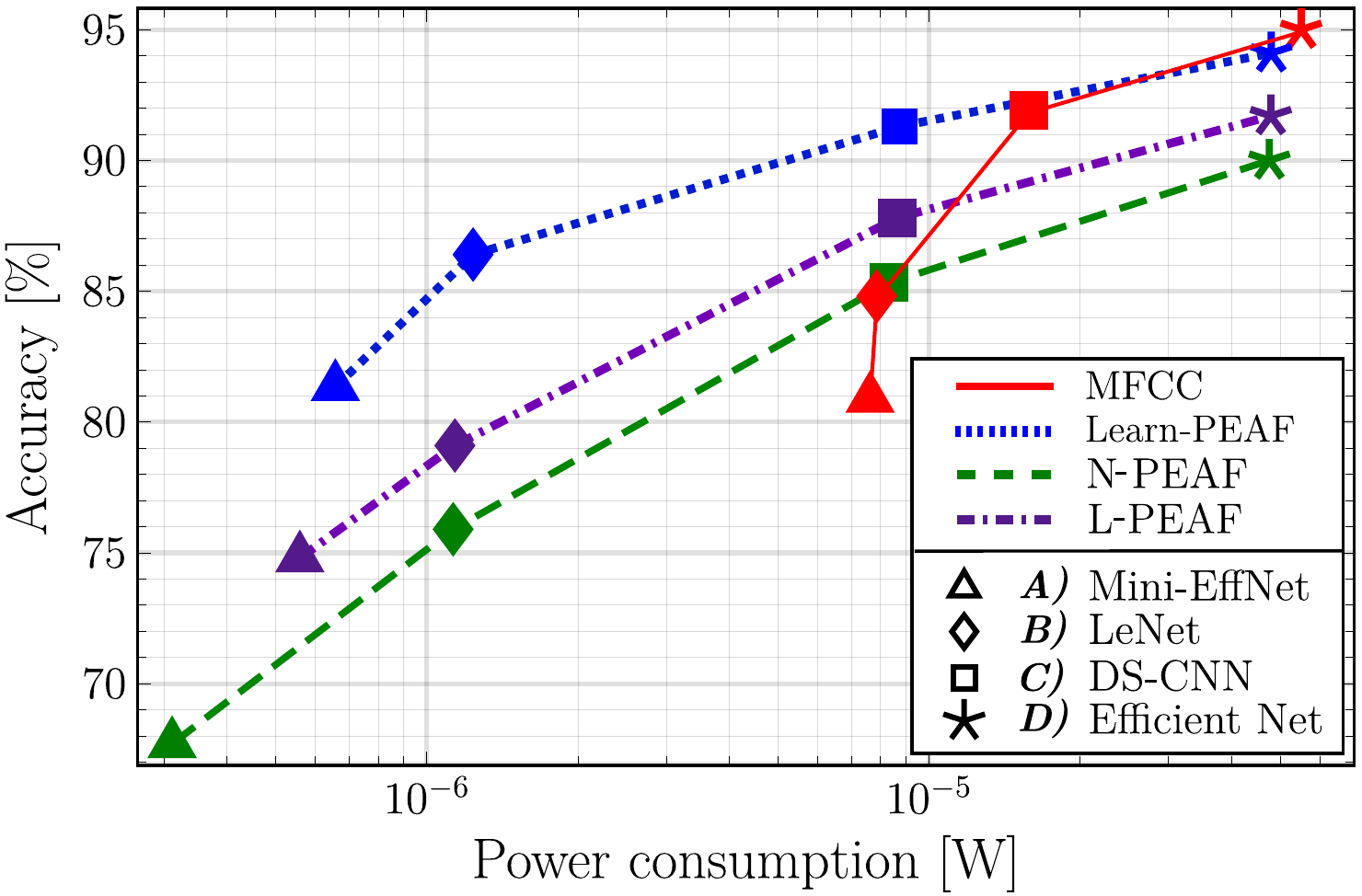}
\caption{Features comparison on KWS.}
\label{fig:KWS_Res}
\end{figure}

%To compare the features on a standard dataset and a on a more complex task the Speech Command dataset V2~\cite{warden2018speech} was used, with all the 35 different classes and no added noise.
%The results for Keyword spotting are summarised in Fig.~\ref{fig:KWS_Res}.
We compared digital MFCCs, and analog PEAF variants on the Speech Command dataset V2~\cite{warden2018speech}, with all the 35 different classes and no added noise. Fig.~\ref{fig:KWS_Res} summarizes obtained results. 
% The parameters of the learnable PEAF are fitted with LeNet Classifier \textbf{(B)} (Sec. \ref{sec:CNN}).

\subsection{Feature comparison for WWD}
\label{sec:WWD}

%We adopted the experimental setup from the Howl work~\cite{tang2020howl} including the code provided in ~\cite{git-howl}. The dataset consisted of multiple positive wake words ``Hey Firefox`` and Common Voice dataset~\cite{ardila2020common} prompts as negative background data. Three hundred fifty actual recordings of ``Hey Firefox`` were also provided. The noise was added from the MS-Noise dataset~\cite{MS-noise} at 10dB SNR. The final dataset (training/validation sets) contained 10\% positive examples (2700 generated and 300 actual recordings) with a total of 30,000 audio samples of 2.5 seconds. 
% Random shifting in the time axis was also added as data augmentation to add even more randomness and prevent boundary conditions influence.
%The testing set was composed of 50 actual recordings of ``Hey Firefox ``, 500 positive generated recordings given by \cite{git-howl}, 3500 background noise, and 3500 negative speech samples, both randomly shifted in time. Fig. \ref{fig:WWD_ROC} shows the ROC curves of different setups.

Evaluating PEAF performance on the second task is two-fold. First, WWD runs on many battery-operated devices, and its power efficiency is critical. Second, Learn-PEAF was fit on the KWS data, and we were interested in evaluating their cross-dataset performance on the different, wake word data.

For WWD, we adopted the open-source experimental setup Howl from Mozilla~\cite{tang2020howl} including the code provided in ~\cite{git-howl}. The dataset consisted of multiple positive wake words ``Hey Firefox`` and Common Voice dataset~\cite{ardila2020common} prompts as negative background data. Three hundred fifty actual recordings of ``Hey Firefox`` were also provided. The noise was added from the MS-Noise dataset~\cite{MS-noise} at 10dB SNR. The final dataset (training/validation sets) contained 10\% positive examples (2700 generated and 300 actual recordings) with a total of 30,000 audio samples of 2.5 seconds. The testing set was composed of 50 actual recordings of ``Hey Firefox ``, 500 positive generated recordings given by \cite{git-howl}, 3500 background noise, and 3500 negative speech samples, both randomly shifted in time.

\begin{figure}[htb]
  \centering
  \includegraphics[width=0.96\linewidth]{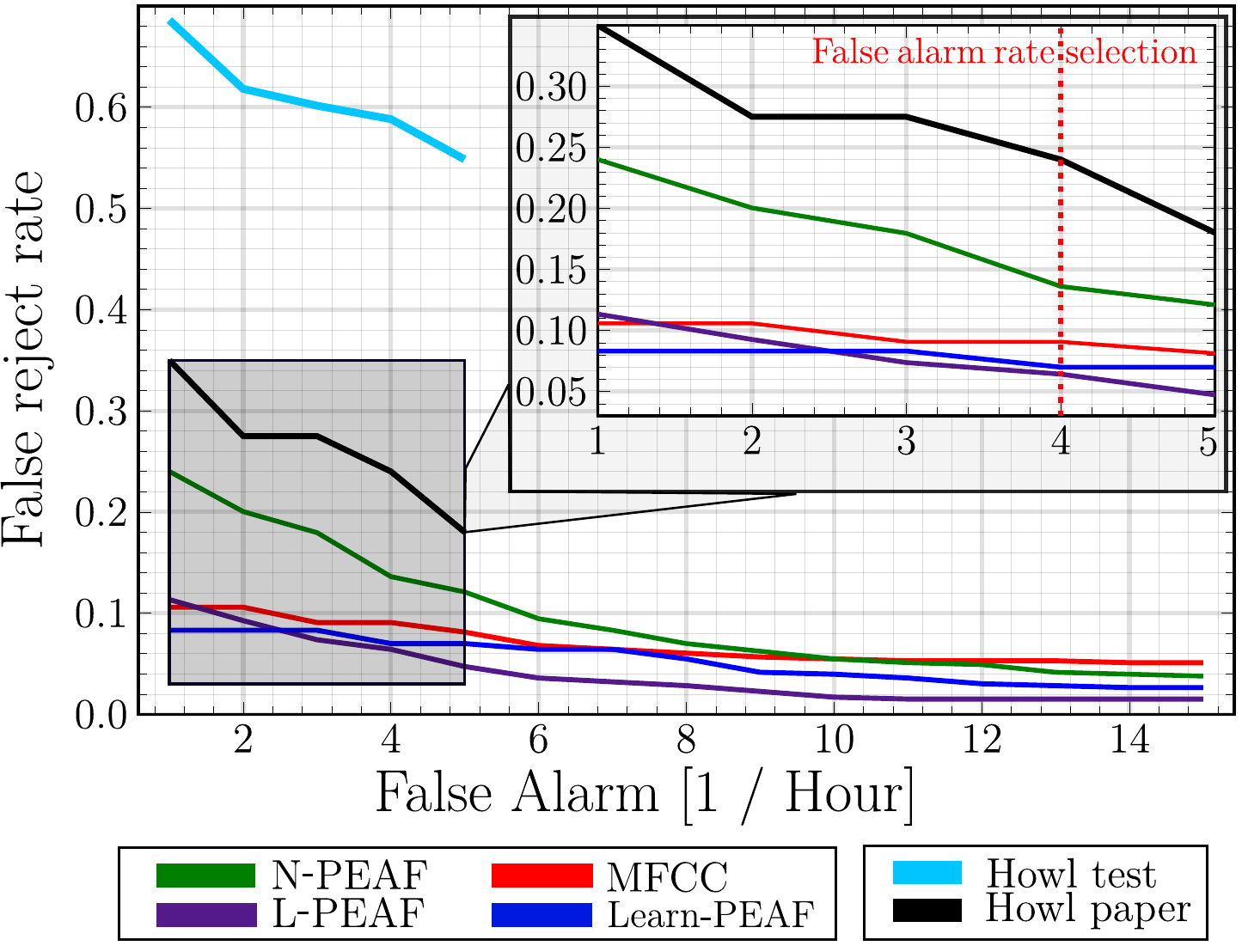}
  \caption{ROC curves for Mini-EfficientNet \textbf{(A)} on WWD.}
  \label{fig:WWD_ROC}
\end{figure}

\begin{figure}[htb]
  \centering
  \includegraphics[width=0.98\linewidth]{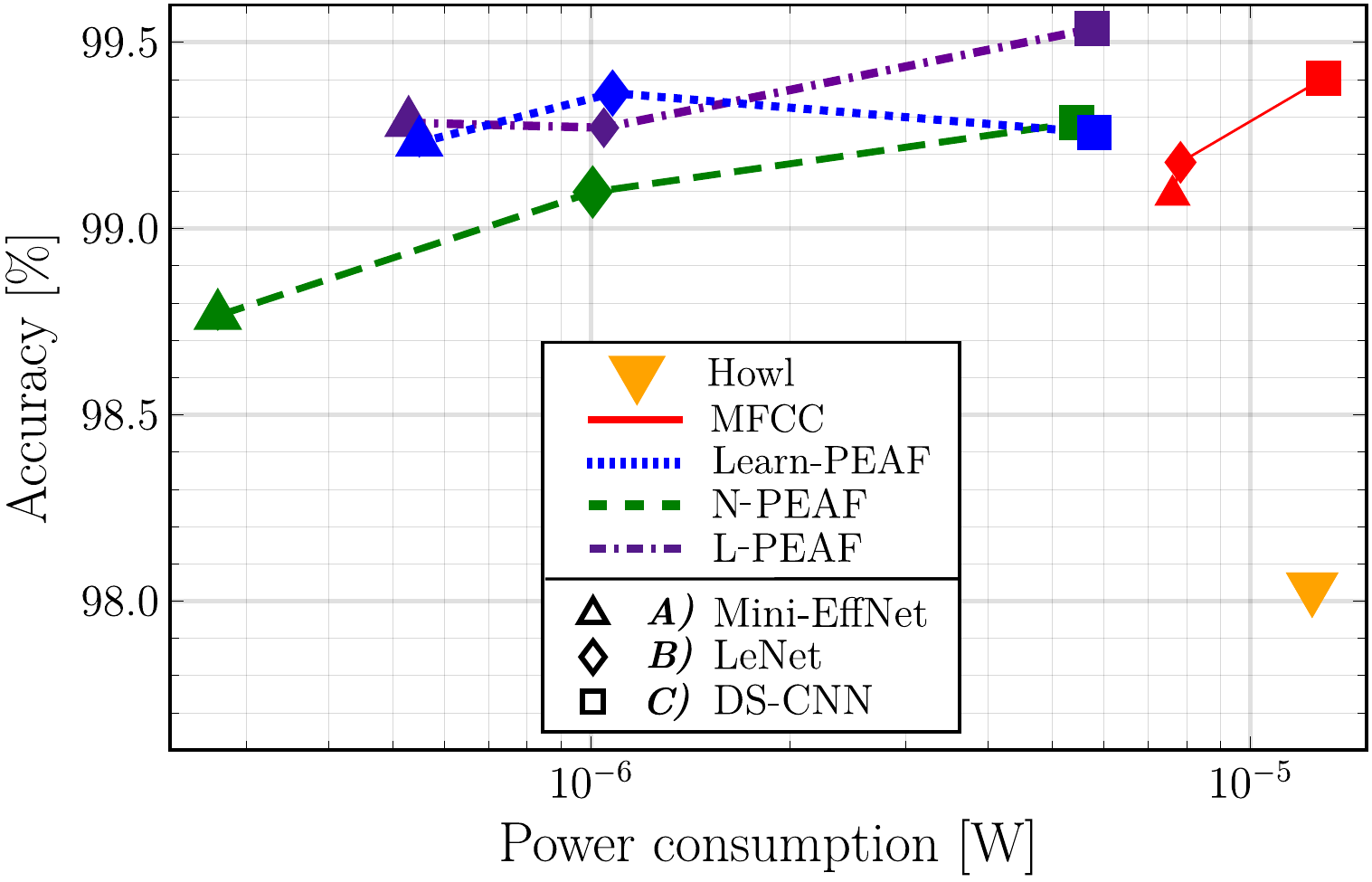}
\caption{Features comparison for WWD.}
\label{fig:WWD_energy}
\end{figure}

Fig. \ref{fig:WWD_ROC} shows the ROC curves of different setups. Howl's performance was evaluated using the pre-trained model (\textit{Res-8 small}) provided by the author~\cite{git-howl}. Fig.~\ref{fig:WWD_energy} shows the accuracy vs. power consumption. To get an accuracy value the false alarm rate was set to 4 as done in Howl.
%The energy consumption of the Howl system was estimated using the MACs of its classifier \textit{Res-8 small}. 

%The ROC curve from Howl ~\cite{tang2020howl} is added even if the dataset is not exactly the same, they have more negative examples. Indeed we were limited by computation time of the analog features. 

\section{Discussion}\label{sec:disc}

Analyzing Fig. \ref{fig:inf_leaf} allows understanding of information bottlenecks. The pattern is similar for all the features; the first step extracts as much information as possible. The dimensionality of the features is then reduced, inducing a reduction in information. 
The MFCC loses a bit of information in the last step. Indeed there are twenty Mel-Frequency bands and only ten final coefficients after DCT. LEAF sees its information increase in mean and variance with the use of PCEN. The output entropy of the MFCC is higher than LEAF, with LEAF having a higher variance. Knowing that the MFCC and LEAF have the same performance on KWS ~\cite{leaf}, it could be a testament to how the variance in entropy is primordial or an issue with our experimental, theoretical framework. The analog features lose the most information with the spikes computation in the IAF and when counting the spikes.
Table \ref{tab:inf_test} shows that all the tested replacements increase the accuracy by a significant amount. Optimizing the parameters of the BPF in the original analog extractor allows for significant and easy to implement improvement. The fully improved Learn-PEAF gains 7.3\% in accuracy, surpassing the MFCC. The other comparison points added on Fig. \ref{fig:KWS_Res} and \ref{fig:WWD_energy} also show a clear improvement. The increase is clear on KWS and WWD with LeNet \textbf{(B)}. Overall this information theory method allowed to get insight into the features and target the computation steps that needed change. A more in depth analysis is required to make it a stable tool. 

The results on keyword spotting (Fig.\ref{fig:KWS_Res}) indicate that L-PEAF performs worse by about 2 to 5\% for all classifiers compared to the digital features. For large classifiers (\textbf{C},\textbf{D}), the energy consumption of the classifier dominates the total power consumption, and there is no reason to choose L-PEAF or N-PEAF over MFCCs. The accuracy gap increases with the smaller neural network (\textbf{A},\textbf{B}) for L-PEAF and N-PEAF. Power consumption is around seven times better for all PEAF with LeNet \textbf{(B)} and more than thirteen times greater for the Mini-EfficientNet \textbf{(A)}. Learn-PEAF seems to favor the small classifiers (\textbf{A,B}) with the improvement of more than 5 \% accuracy compared to L-PEAF. The larger neural networks (\textbf{C,D}) show only a 2 \% accuracy improvement. Overall, Learn-PEAF outperforms the MFCC in the low power domain on both axes.

Fig.\ref{fig:WWD_ROC} shows that L-PEAF has the overall best ROC curves, followed by Learn-PEAF closely. Evaluated Howl's pre-trained system demonstrated a non-competitive ROC curve. A difference in pre-processing could explain it. The weights were used as-is and not retrained. The reported Howl ROC curves are in the same magnitude as our tested results. Apart from the results from N-PEAF, all the ROC curves are close. 
The accuracy results on Fig.\ref{fig:WWD_energy} demonstrate a potential usage of L-PEAF and Learn-PEAF, with results better than the MFCC. The gain is also evident in energy efficiency for all PEAFs, with a clear energy gap between analog and digital features. N-PEAF could be an alternative for the most power efficient WWD application consuming two times less energy compared to L-PEAF  with a 1 \% accuracy loss. Learn-PEAF outperform all the other features only with LeNet \textbf{(B)}. As it was trained with this classifier on keyword spotting, it is interesting to note that it can transfer some of its knowledge to another task. It could be improved further by training its parameters on the WWD task.

One limitation of our comparison is that Learn-PEAF uses a simplified version of the IAF encoder. It was done to increase computing time dramatically and allow the training of the extractor's parameters. Both L-PEAF and N-PEAF are also simulated numerically, no recording of the real analog chips were taken. This was done in \cite{linear, nonlinear}; though only for VAD, the simulation proved to be accurate enough.

\section{Conclusions}

%Milos
%This paper compared different feature extraction methods and showed the potential of the analog features. The WWD task showed that the analog feature extractor could outperform the digital ones, with an order of magnitude lower power consumption. We then identified the bottleneck in the analog feature extractor's computation pipeline using an information theory framework, which was validated with a 7\% accuracy improvement on keyword spotting by replacing the bottleneck stages with some building blocks from LEAF. Future works include adapting the processing in the modified stages to feasible integrated electronics implementation and exploring the possibility of directly using spiking neural networks as the classifier after IAF for accuracy improvement.

This paper has presented linear, non-linear, and learnable power-efficient acoustic analog features, and evaluated them on two speech processing tasks demanded in many battery-operated devices: WWD and KWS. The WWD task showed that the analog features outperform the digital ones, with an order of magnitude lower power consumption and competitive classification accuracies. 

We argue that the novel information theory-based method introduced in this paper could also be applied for other feature analysis. We have used it to identify the information bottleneck in L-PEAF calculation and, inspired by LEAF, devised a novel learn-PEAF. Learn-PEAF improved the KWS accuracy up to 7\% compared to L-PEAF, outperforming MFCCs with less power consumption.

Future works include adapting the processing in the modified stages to feasible integrated electronics implementation and exploring the possibility of directly using spiking neural networks as the classifier after IAF for accuracy improvement.

\newpage
\bibliographystyle{IEEEtran}
\balance
\bibliography{mybib}

\begin{thebibliography}{10}
\providecommand{\url}[1]{#1}
\csname url@samestyle\endcsname
\providecommand{\newblock}{\relax}
\providecommand{\bibinfo}[2]{#2}
\providecommand{\BIBentrySTDinterwordspacing}{\spaceskip=0pt\relax}
\providecommand{\BIBentryALTinterwordstretchfactor}{4}
\providecommand{\BIBentryALTinterwordspacing}{\spaceskip=\fontdimen2\font plus
\BIBentryALTinterwordstretchfactor\fontdimen3\font minus
  \fontdimen4\font\relax}
\providecommand{\BIBforeignlanguage}[2]{{%
\expandafter\ifx\csname l@#1\endcsname\relax
\typeout{** WARNING: IEEEtran.bst: No hyphenation pattern has been}%
\typeout{** loaded for the language `#1'. Using the pattern for}%
\typeout{** the default language instead.}%
\else
\language=\csname l@#1\endcsname
\fi
#2}}
\providecommand{\BIBdecl}{\relax}
\BIBdecl

\bibitem{hosticka1985performance}
B.~J. Hosticka, ``Performance comparison of analog and digital circuits,''
  \emph{Proceedings of the IEEE}, vol.~73, no.~1, pp. 25--29, 1985.

\bibitem{linear}
M.~Yang, C.-H. Yeh, Y.~Zhou, J.~P. Cerqueira, A.~A. Lazar, and M.~Seok, ``A
  1$\mu$w voice activity detector using analog feature extraction and digital
  deep neural network,'' in \emph{2018 IEEE International Solid-State Circuits
  Conference (ISSCC)}, 2018, pp. 346--347.

\bibitem{nonlinear}
M.~Yang, H.~Liu, W.~Shan, J.~Zhang, I.~Kiselev, S.~J. Kim, C.~Enz, and M.~Seok,
  ``{Nanowatt Acoustic Inference Sensing Exploiting Nonlinear Analog Feature
  Extraction},'' \emph{IEEE Journal of Solid-State Circuits}, vol.~56, no.~10,
  pp. 3123--3133, 2021.

\bibitem{raychowdhury20132}
A.~Raychowdhury, C.~Tokunaga, W.~Beltman, M.~Deisher, J.~W. Tschanz, and V.~De,
  ``A 2.3 nj/frame voice activity detector-based audio front-end for
  context-aware system-on-chip applications in 32-nm cmos,'' \emph{IEEE journal
  of solid-state circuits}, vol.~48, no.~8, pp. 1963--1969, 2013.

\bibitem{mfcc-energy}
W.~Shan, M.~Yang, T.~Wang, Y.~Lu, H.~Cai, L.~Zhu, J.~Xu, C.~Wu, L.~Shi, and
  J.~Yang, ``A 510-nw wake-up keyword-spotting chip using serial-fft-based mfcc
  and binarized depthwise separable cnn in 28-nm cmos,'' \emph{IEEE Journal of
  Solid-State Circuits}, vol.~56, no.~1, pp. 151--164, 2021.

\bibitem{chandrakumar201815}
H.~Chandrakumar and D.~Markovi{\'c}, ``{A 15.2-ENOB 5-kHz BW 4.5-$\mu $ W
  Chopped CT $\Delta\Sigma $-ADC for Artifact-Tolerant Neural Recording Front
  Ends},'' \emph{IEEE Journal of Solid-State Circuits}, vol.~53, no.~12, pp.
  3470--3483, 2018.

\bibitem{Dellaferrera_2020}
G.~Dellaferrera, F.~Martinelli, and M.~Cernak, ``{A Bin Encoding Training of a
  Spiking Neural Network Based Voice Activity Detection},'' in \emph{{ICASSP
  2020 IEEE International Conference on Acoustics, Speech and Signal Processing
  (ICASSP)}}.\hskip 1em plus 0.5em minus 0.4em\relax IEEE, May 2020, pp.
  3207--3211.

\bibitem{martinelli2020spiking}
F.~Martinelli, G.~Dellaferrera, P.~Mainar, and M.~Cernak, ``Spiking neural
  networks trained with backpropagation for low power neuromorphic
  implementation of voice activity detection,'' in \emph{{ICASSP 2020 IEEE
  International Conference on Acoustics, Speech and Signal Processing
  (ICASSP)}}.\hskip 1em plus 0.5em minus 0.4em\relax IEEE, May 2020, pp.
  8544--8548.

\bibitem{leaf}
N.~Zeghidour, O.~Teboul, F.~de~Chaumont~Quitry, and M.~Tagliasacchi, ``Leaf: A
  learnable frontend for audio classification,'' 2021.

\bibitem{Eff-V2}
\BIBentryALTinterwordspacing
M.~Tan and Q.~V. Le, ``Efficientnetv2: Smaller models and faster training,''
  \emph{CoRR}, vol. abs/2104.00298, 2021. [Online]. Available:
  \url{https://arxiv.org/abs/2104.00298}
\BIBentrySTDinterwordspacing

\bibitem{726791}
Y.~Lecun, L.~Bottou, Y.~Bengio, and P.~Haffner, ``Gradient-based learning
  applied to document recognition,'' \emph{Proceedings of the IEEE}, vol.~86,
  no.~11, pp. 2278--2324, 1998.

\bibitem{howard2019searching}
A.~Howard, M.~Sandler, G.~Chu, L.-C. Chen, B.~Chen, M.~Tan, W.~Wang, Y.~Zhu,
  R.~Pang, V.~Vasudevan, Q.~V. Le, and H.~Adam, ``{Searching for
  MobileNetV3},'' in \emph{{Proceedings of the IEEE/CVF International
  Conference on Computer Vision}}, 2019, pp. 1314--1324.

\bibitem{tan2020efficientnet}
M.~Tan and Q.~Le, ``{Efficientnet: Rethinking model scaling for convolutional
  neural networks},'' in \emph{International conference on machine
  learning}.\hskip 1em plus 0.5em minus 0.4em\relax PMLR, 2019, pp. 6105--6114.

\bibitem{kingma2014adam}
D.~P. Kingma and J.~Ba, ``Adam: A method for stochastic optimization,''
  \emph{arXiv preprint arXiv:1412.6980}, 2014.

\bibitem{Spec-augment}
D.~S. Park, W.~Chan, Y.~Zhang, C.-C. Chiu, B.~Zoph, E.~D. Cubuk, and Q.~V. Le,
  ``{SpecAugment: A Simple Data Augmentation Method for Automatic Speech
  Recognition},'' in \emph{Proc. Interspeech 2019}, 2019, pp. 2613--2617.

\bibitem{tu202228nm}
F.~Tu, Y.~Wang, Z.~Wu, L.~Liang, Y.~Ding, B.~Kim, L.~Liu, S.~Wei, Y.~Xie, and
  S.~Yin, ``{A 28nm 29.2 TFLOPS/W BF16 and 36.5 TOPS/W INT8 Reconfigurable
  Digital CIM Processor with Unified FP/INT Pipeline and Bitwise In-Memory
  Booth Multiplication for Cloud Deep Learning Acceleration},'' in \emph{{2022
  IEEE International Solid-State Circuits Conference (ISSCC)}}, vol.~65.\hskip
  1em plus 0.5em minus 0.4em\relax IEEE, 2022, pp. 1--3.

\bibitem{Shanon_entropy}
C.~E. Shannon, ``A mathematical theory of communication,'' \emph{The Bell
  system technical journal}, vol.~27, no.~3, pp. 379--423, 1948.

\bibitem{Bergsma-thesis}
B.~Bergsma, ``{Power efficient acoustic feature representation for IoTs},''
  Master's thesis, École Polytechnique Fédérale de Lausanne, 2021.

\bibitem{warden2018speech}
P.~Warden, ``{Speech Commands: A Dataset for Limited-Vocabulary Speech
  Recognition},'' \emph{arXiv preprint arXiv:1804.03209}, 2018.

\bibitem{tang2020howl}
\BIBentryALTinterwordspacing
R.~Tang, J.~Lee, A.~Razi, J.~Cambre, I.~Bicking, J.~Kaye, and J.~Lin, ``Howl: A
  deployed, open-source wake word detection system,'' in \emph{Proceedings of
  Second Workshop for NLP Open Source Software (NLP-OSS)}.\hskip 1em plus 0.5em
  minus 0.4em\relax Association for Computational Linguistics, Nov. 2020, pp.
  61--65. [Online]. Available:
  \url{https://www.aclweb.org/anthology/2020.nlposs-1.9}
\BIBentrySTDinterwordspacing

\bibitem{git-howl}
------, ``The github page of howl.'' \\
  \url{https://github.com/castorini/howl}.

\bibitem{ardila2020common}
\BIBentryALTinterwordspacing
R.~Ardila, M.~Branson, K.~Davis, M.~Kohler, J.~Meyer, M.~Henretty, R.~Morais,
  L.~Saunders, F.~Tyers, and G.~Weber, ``\BIBforeignlanguage{English}{Common
  voice: A massively-multilingual speech corpus},'' in
  \emph{\BIBforeignlanguage{English}{Proceedings of the 12th Language Resources
  and Evaluation Conference}}.\hskip 1em plus 0.5em minus 0.4em\relax
  Marseille, France: European Language Resources Association, May 2020, pp.
  4218--4222. [Online]. Available:
  \url{https://aclanthology.org/2020.lrec-1.520}
\BIBentrySTDinterwordspacing

\bibitem{MS-noise}
C.~K. Reddy, E.~Beyrami, J.~Pool, R.~Cutler, S.~Srinivasan, and J.~Gehrke, ``{A
  Scalable Noisy Speech Dataset and Online Subjective Test Framework},'' in
  \emph{Proc. Interspeech 2019}, 2019, pp. 1816--1820.

\end{thebibliography}

\end{document}